\documentstyle[12pt]{article}
\def\binom#1#2{{#1\choose#2}}
\def\ds{\displaystyle}
\evensidemargin\oddsidemargin
\begin{document}
\newtheorem{theorem}{theorem}
\newtheorem{proposition}{proposition}
\newtheorem{definition}{definition}
\newtheorem{lemma}{lemma}
\newtheorem{notation}{notation}
\begin{flushright}
SISSA-ISAS 92/98/FM-EP\\
hep-th/9808170\\
\end{flushright}
\vspace{1cm}

\begin{center}
{\large {Explicit Computations for the Intersection Numbers on
Grassmannians, and on the Space of Holomorphic Maps from $CP^1$ into
$G_r(C^n)$}}   

\vspace{1cm}
Noureddine Chair\footnote{e-mail:n.chair@rocketmail.com}\\
{\it Faculty of Art and Sciences, Physics Department,}\\
{\it Al al-Bayt University Mafraq, Jordan}\\ 
{\it and}\\
{\it SISSA/ISAS Via Beirut 2, 34014 Trieste, Italy}

\end{center}
\vspace{1.5cm}

\begin{abstract} 
We derive some explicit expressions for correlators on Grassmannian
$G_r(C^n)$ as well as on the moduli space of holomorphic maps, of a
fixed degree $d$, from sphere into the Grassmannian. Correlators
obtained on the Grassmannain are a first step generalization of the
Schubert formula for the self-intersection. The intersection numbers
on the moduli space for $r=2,3$ are given explicitly by two closed
formulas, when $r=2$ the intersection numbers, are found to generate
the alternate Fibonacci numbers, the Pell numbers and in general a
random walk of a particle on a line with absorbing barriers. For $r=3$
the intersection numbers form a well organized pattern.
\end{abstract} 
\vfill 
\eject
 
\section{Introduction} 

The classical Schubert calculus computes intersection numbers on
Grassmannians $G_r(C^n)$ of complex r-planes in $C^n$, by using the
Giambelli and Pieri formula \cite {H,Gr,F}. It is due primarily to
Schubert more than a hundred years ago, to obtain the number,
\begin{eqnarray}
\int_{G_r(C^n)} x_1^{r(n-r)} = \frac{1!2!3! \cdots (r-2)! (r-1)!
(r(n-r))!}{(n-r)! (n-r+1)! \cdots (n-1)!} \nonumber \, ,
\end{eqnarray} 
known as the degree of the Grassmannian or the self-intersection of
the first Chern class $x_1$, of the r-plane bundle $Q$ on
$G_r(C^n)$. Geometrically speeking this number corresponds to the
number of $(r-1)$-planes in $CP^{n-1}$ meeting $r(n-r)$ general
$(n-r-1)$-planes, in particular for $r=2$ there are two lines meeting
4-given lines in $CP^3$.

Our goals in this paper are two-fold, first we would like to extend
the above formula to other correlators that are products of Chern
classes $x_i$, $r \leq i \leq r$. In this direction, we use the
pairing residue formula that computes the correlatores in topological
Landau-Ginzburg Theories \cite {V} and the explicit formula for the
potential $W(x_i)$ that generate the cohomology ring \cite {C}, to do
some explicit computations on the Grassmannian. The different
intersection numbers obtained, show a certain pattern amongst
themselves and is formulated in a proposition which in turn lead to
the closed formula for $\int_{G_r(C^n)} x_1^{r(n-r)-rk_{r-1}}
x_r^{k_{r-1}}$.

The second goal is to carry out similar computations on the space of
holomorphic maps of a fixed degree $d$, from a Riemann surface of
genus zero ($CP^1$) into the Grassmannian $G_r(C^n)$. Formally both
computations use the same formula \cite {V,In}, the difference between
the two cases is that the potential in the second case is a deformed
potential $\widetilde {W} (x_i)$ and is connected to the previous
potential by $\widetilde {W} (x_i)= W(x)+(-1)^r q x_1$ \cite
{V3}. This potential reproduces the quantum cohomology ring of the
Grassmannian \cite {W2}. The concept of deformation of the cohomology ring
was first observed in \cite {Wi}, in connection with the $CP^1$
model. On the space of holomorphic maps from $CP^1$ into $G_r(C^n)$, we
have two closed formulas for $r=2$ and $3$ for any $n$. When $r=2$,
the intersection numbers generate well known numbers like the
Fibonacci numbers and the Pell numbers for $n=5,6$ respectively ,and
when $n \geq 7$, the intersection numbers generates a random walk of a
particle on a line with absorbing barriers \cite {Sl,E}.Our closed
formula for the intersection numbers on the space of holomorphic maps
into $G_3(C^n)$, when restricted to constant maps, gives all the
intersection numbers on $G_3(C^n)$. Some intersection numbers on this
space were computed when $n=6$ and for degree one and two. We find
that these numbers organize themselves in an ordered pattern, it seems
that the intersection numbers on this moduli space is a zoo of
interesting numbers. This fact is already presented on the
Grassmannian $G_r(C^n)$; if we set $ N=n-r $ in the Schubert formula,
we obtain the generating functions for the $r$-dimensional Catalan
numbers \cite{Sl}. When $r=2$, we obtain the ordinary Catalan numbers
$(2N)!/(N+1)!N!$.

This paper is organized as follows: in section 2 after a brief account
of the cohomology ring of the Grassmannian, the pairing residue
formula which computes correlators in $N=2$ topological Landau
theories, and fixing our notations, we compute some intersection
number on the Grassmannian $G_r(C^n)$ from which we obtaine closed
formula for the correlators. Section 3 and 4 will be devoted to computations
of correlators on the space of holomorphic maps from $CP^1$ into
$G_r(C^n)$, in which we find connections between intersection numbers,
Fibonacci numbers, Pell numbers and the random walk. Our conclusions
are given in section 5.

\section{Intersection Numbers on a Grassmannian}

In this section we shall first recall briefly the definition of the
cohomology ring of the Grassmannian $G_r(C^n)$ in the Landau-Ginzburg
formulation \cite{V2,Wi,G} and the pairing residue formula of the N=2
topological Landau-Ginzburg model that computes the correlators
\cite{V}.  We then use the specialized form of ``the pairing residue
formula'' and the explicit expression for the Landau-Ginzburg
potential in terms of the generators $x_i(1 \leq i \leq r)$ of the
chomology ring of the Grassmannian \cite{C} to compute some
intersection numbers. The intersection numbers computed are exactly
those obtained using the Schubert Calculus \cite{Gr,F}. These
computations show that the two-point function $\left<x_1^{\alpha_1}
x_2^{\alpha_2 }\right>$ on $G_2(C^{n+1})$ is equal to the two point
function $\left<x_1^{\alpha_1} x_2^{\alpha_2 -1}\right>$ on
$G_2(C^n)$. In general the r-point functions on $G_r(C^{n+1})$ and
$G_r(C^{n})$ are related in the same way.  This fact will be proved in
the proposition below and as a consequence we obtain an explicit
expression for the two point functions on $G_r(C^n)$ involving the
Chern classes $x_1$ and $x_r$.

The cohomology ring of the complex Grassmannian manifold, denoted by
$H^*(G_r(C^n))$ is a truncated polynomial ring in several variables
\cite {BT} given by
\begin{equation}
H^*(G_r(C^n))\cong C[x_1,\cdots,x_r,y_1,\cdots ,y_{n-r}]/I\,,
\label{alpha}
\end{equation}
where $x_i=c_i(Q)$ (for $1\leq i\leq r$) are the Chern classes of the
quotient bundle $Q$ of rank $r$, {\it i.e.}, $x_i\in H^{2i}(G_r(C^n))$
and $y_j=c_j(S)$ (for $1\leq j\leq n-r)$ are the Chern classes of the
universal bundle $S$ of rank $n-r$.  The ideal $I$ in
$C[x_1,\cdots,x_r,y_1,\cdots,y_{n-r}]$ is given by
\begin{equation}
(1+x_1+x_2+\cdots+x_r)(1+y_1+y_2+\cdots +y_{n-r})=1\,,
\label{relxy}
\end{equation}
which is the consequence  of the tautological sequence on $G_r(C^n)$ 
$$
0\longrightarrow S\longrightarrow V\longrightarrow Q\longrightarrow
0\,, $$
where $ V=G_r(C^n)\times C^n$. By using equation (\ref{relxy}), one
may rewrite $H^*(G_r(C^n))$ as
\begin{equation}
H^*(G(C^n))\cong C[x_1,\cdots, x_r]/y_j\,,
\label{c}
\end{equation}
where $y_j$ are expressed in terms of $x_i$, and $y_j=0$ for
$n-r+1\leq j\leq n$, and $x_0=y_0=1$. The classes $ y_j $ can be
written inductively as a function of $x_1,\cdots,x_r$ via
\begin{equation}
y_j=-x_1y_{j-1}-\cdots-x_{j-1}y_1-x_j\,,\quad {\rm for}\,\,
j=1,\cdots, n-r\,.
\label{e}
\end{equation}   
In the Landau-Ginzburg formulation, the potential that generates the 
cohomolgy ring of the grassmannian  \cite{V2,G,Wi}, is given by
\begin{equation}
W_{n+1}(x_1,\cdots ,x_r)=\sum_{i=1}^{r} \frac{q_i^{n+1}}{n+1}\,,  
\end{equation}  
where, $x_i$ and $q_i$ are related by
\begin{equation}
x_i=\sum_{1\leq l_1 < l_2 \cdots < l_i \leq r} q_{l_1} q_{l_2} \cdots
q_{l_i} \,.
\end{equation}
The cohomology ring of the grassmannian is then given by
\begin{equation}
\frac{\partial W_{n+1}}{\partial x_i}=(-1)^n y_{n+1-i}, \quad {\rm
for} 1 \leq i \leq r\,,
\end{equation}
implying that $d_iW_{n+1}=0$, for $i=1,\cdots, r$. In terms of the
$x_i's$, \cite{C} the explicit formulas for the $y_j's$ and the
cohomology potential $W(x_1,....x_r)$ are,
\begin{eqnarray}
{\ds y_j }&= &  (-1)^j {\ds \sum_{k_1=0}^{\left[\frac{j}{2}\right]}
\cdots \sum_{k_{r-1}=0}^{\left[\frac{j}{r}\right]}
\frac{ (-1)^{k_1+2k_2+\cdots+(r-1)k_{r-1}}}{k_1!\cdots
k_{r-1}!} \times } \nonumber \\ {}\nonumber\\
& & \quad{\ds \frac{\left(j- \sum_{l=1}^{r - 1} l k_l \right)!}
{\left(j - \sum_{l=2}^{r} l k_{l-1} \right)!}
x_1^{j-2k_1-\cdots rk_{r-1}} x_2^{k_1}\cdots x_r^{k_{r-1}} }\,.
\label{k}
\end{eqnarray}
\begin{eqnarray}
{\ds W_{n+1}(x_1,\cdots ,x_r) }&=& {\ds \sum_{k_1=0}^{\left
[\frac{n+1}{2}\right]}\cdots
\sum_{k_{r-1}=0}^{\left[\frac{n+1}{r}\right]}
\frac{ (-1)^{k_1+2k_2+\cdots+(r-1)k_{r-1}}}{k_1!\cdots
k_{r-1}!} \times } \nonumber \\ 
{}\nonumber\\
&& {\ds \frac{\left(n- \sum_{j=1}^{r - 1} j k_j \right)!}
{\left(n+1 - \sum_{j=2}^{r} j k_{j-1} \right)!}
x_1^{n+1-2k_1-\cdots rk_{r-1}} x_2^{k_1}\cdots x_r^{k_{r-1}} }.
\label{a}
\end{eqnarray}
The self-intersection numbers $\left< x_1^{r(n-r)}\right>$, and other
correlation functions on the Grassmannian $G_r(C^n)$ of products of
monomials in the cohomology classes $x_i(1 \leq i \leq r)$ such that
the total power of this product is the dimension of $G_r(C)$,i.e.,
$r(n-r)$, may be computed using the residue pairing formula
\cite{V}. This formula computes the correlators in the topological
Landau-Ginzburg theories, which for genus zero, reads
\begin{eqnarray}
{\ds\left<\prod_{i=1}^{N} F_i(x_j)\right>}&=&{\ds (-1)^{{N(N-1)}/2} 
\sum_{dW=0}\frac{\prod_{i=1}^{N} F_i(x_j)}{H} } \nonumber \\ 
{} \nonumber \\
&=&{\ds (-1)^{N(N-1)/2} \frac{1}{{(2 \pi i)}^N} \oint \cdots \oint 
\frac{dx_1 \cdots dx_N \prod_{i=1}^{N} F_i (x_j) }{\partial_1 W 
\cdots \partial_N W}}\,,
\end{eqnarray} 
where $F_i(x_j)$ are polynomials in the superfields $x_i$, $H=det
(\partial_i \partial_j W)$ is the hessian and the summation on the
right hand side in the first expantion is over the critical points of
$W$. In this section, the maps from sphere into $G_r(C^n)$ are considered
constant, i.e., the moduli space of instantons is nothing but the
Grassmannian $G_r(C^n)$ itself, and the correlators are the
intersections of the cycles over $G_r(C^n)$. Therefore the residue
pairing formula reads\footnote{This is a natural parametrization for
the powers of $x_i's$ since the total power sums up to $r(n-r)$, 
otherwise the correlators vanish.}
\begin{equation}
{\ds \left< x_1^{n+1-2k_1- \cdots rk_{r-1}} x_2^{k_1}  \cdots
x_r^{k_{r-1}} \right>} = 
{\ds \frac{(-1)^{r(r-1)/2}}{(2 \pi i)^r} \oint \cdots \oint
\frac{x_1^{n+1-2k_1-\cdots rk_{r-1}} x_2^{k_1}\cdots x_r^{k_{r-1}}}
{{\partial_1 W \cdots \partial_r W}} }\,  
\end{equation}

The closed form for these correlators is ,in general, not known,
except for the self-intersection $\left< x_1^{r(n-r)}\right>$, which
was given by the Schubert calculus \cite{H,Gr,F} (also called the
degree of the Grassmannian) and has the following expression
\begin{equation} 
{\ds \left<x_1^{r(n-r)} \right> }= {\ds (r(n-r))! \prod_{\ell
=0}^{r-1} \frac{\ell!}{(n-r+\ell)!} } \, . 
\end{equation}   
In particular for $G_2(C^4)$ the self-intersection
$\left<x_1^4\right>$, is 2 which is the number of lines meeting 4
given lines in $CP^1$, and in general the right hand side of the above
equation gives the number of $(r-1)$-planes meeting $N=r(n-r)$ given
$(n-(r+1))$-planes in general position in $CP^{n-1}$.  The simplest
non-trivial Grassmannian for which the residue pairing formula can
be used is $G_2(C^4)$. Here the potential that generates the
cohomology ring (intersection ring) $H^*(G_2(C^4))$, is
$W(x_1,x_2)=\frac{1}{5} x_1^5 -x_1^3 x_2 + x_1 x_2^2$ and the possible
correlators are $<x_1^{4-2k} x_2^k>$ where $0\leq k \leq 2$. Applying
the residue pairing formula we have,
\begin{equation} 
{\ds \left< x_1^{4-2k} x_2^k \right>}={\ds -\frac{1}{(2 \pi i)^2} 
\oint\oint\frac{x_1^{4-2k} x_2^k dx_1 dx_2}{(x_1^4 -3x_1^2 x_2 +x_2^2)
(-x_1^3+2x_1x_2)} }\,. 
\end{equation} 
Explicit computation for $k=0,1,2$ gives ($ \partial_2 w=0$ for   
$x_2=\frac{1}{2} x_1^2$) $\left<x_1^4 \right>=2$, 
$\left< x_1^2 x_2\right> =1$ and $\left< x_2^2 \right> =1$, which agree
with the Schubert calculus \cite{F}. In the same way, we have computed
the corrlators  $I_k^n:=\left< x_1^{2(n-2)-2k}x_2^k \right>$ for
$n=5,6,7,$ on $G_2(C^n)$ and $I_{k_1 k_2}^n:=\left< x_1^{3(n-3)-2k_1
-3k_2} x_2^{k_1} x_3^{k_2} \right>$ on $G_3(C^n)$ for $n=5,6,7$ and
the results obtained are  indicated in tables 1 and 2. 
  
We have checked our computations using the property, $Res_W(H)=\mu$,
where $\mu$ is the criticality index of $W$ \cite{V}, {\it i.e.},the
dimension of chiral ring $R=\frac{C[x_i]}{dW_i}$. The above
computations on $G_2 (C^n)$ and $G_3 (C^n)$ indicate that we should
have $\left<x_1^{\alpha_1} x_2^{\alpha_2 }\right>_{G_{2}(C^{n+1})} =
\left<x_1^{\alpha_1} x_2^{\alpha_2 -1}\right>_{G_{2}(C^n)}$ ,
$\left<x_1^{\alpha_1} x_2^{\alpha_2}
x_3^{\alpha_3}\right>_{G_{3}(C^{n+1})} = \left<x_1^{\alpha_1}
x_2^{\alpha_2 } x_3^{\alpha_{3}-1} \right>_{G_{3}(C^n)} $ and, in
general, 
$$
\left< x_1^{\alpha_1}x_2^{\alpha_2} \dots
x_r^{\alpha_r}\right>_{G_{r}(C^{n+1})} = \left<
x_1^{\alpha_1}x_2^{\alpha_2} \dots
x_r^{\alpha_{r}-1}\right>_{G_{r}(C^n)}
$$ 
with $ \sum_{i=1}^{r-1}\alpha_i =r(n-r)$. This is indeed the case as
we shall show in the proposition below; but first we need the
following lemma 
\begin{lemma}
Given an inclusion i:$Z \hookrightarrow X$ (non singular subvariety)
with $dim_C X- dim_C Z = r=n-m$ and suppose there exists a complex
vector bundle E on X such that $E |_Z = N_{Z,X}$ (normal bundle of Z
in X) and $\alpha \in H^{2m} (X)= H^{2n-2r} (X)$ then $i^* (\alpha
)=\alpha X_r (E)$.
\end{lemma} 
For simplicity consider the case $G_2(C^n) \hookrightarrow
G_2(C^{n+1})$, and let $^{n+1}Q$ and $^n Q$ denote the quotient
subundles on $G_2(C^{n+1})$ and $G_2(C^n)$ respectively, both of rank
2. Then the induced pullback gives $i^* ({}^{n+1}{Q} )={}^n {Q}$,
furthermore $ {}^{n+1}{Q} |_{G_2(C^n)}=
N_{G_2(C^n),G_2(C^{n+1})}$. The above remarks on the intersection
numbers on $G_2(C^n)$, $G_2(C^{n+1})$ computed by the residue pairing
formula are equivalent to the following proposition:
\begin{proposition} 
The correlators on $G_2(C^{n+1})$ and $G_2(C^n)$ are identical in the
following sense 
\begin{eqnarray} 
\left<x_1 ( {}^{n+1} {Q} )^{2n-4-2k}  x_2 ( {}^{n+1} {Q} )^{k+1}\right> =
 \left<x_1 ( {}^{n} {Q} )^{2n-4-2k}  x_2 ( {}^{n} {Q} )^{k}\right>
\end{eqnarray}
Proof: setting $x_1 ({}^{n} {Q} )^{2n-4-2k} = U, x_1({}^{n+1} {Q}
)^{2n-4-2k} = U^\prime $, $x_2 ( {}^{n} {Q} )^k = V^k$ and $x_2 (
{}^{n+1} {Q} )^k = V^{\prime k}$, applying the above lemma; $i^*(
\alpha )= \alpha x_r ( {}^{n+1} {Q} )$ with $r=2$ then $i^* (U^\prime
V^{\prime k} ) = (U^\prime V^{\prime k} ) V^\prime = U^\prime
V^{\prime {k+1}}$ and by using the homomorphism of the pullback, the
lefthand side is $U.V$, hence the proof of the proposition.
\end{proposition}

The above proposition can be generalized to correlators on $G_r(C^n)$,
namely, we will have the following
\begin{eqnarray}
&&{\ds \left<x_1 ( {}^{n+1} {Q} )^{r(n-r)-2k_1 -3k_2 -\cdots -rk_{r-1}} x_2
( {}^{n+1} {Q} )^{k_1} \cdots x_r
( {}^{n+1} {Q} )^{k_{r-1}} \right> }=  \nonumber \\ 
&&\qquad\qquad\qquad\qquad
{\ds \left<x_1 ( {}^{n} {Q} )^{r(n-r)-2k_1 -3k_2 -\cdots -rk_{r-1}} x_2
({}^{n} {Q} )^{k_1}\cdots x_r ({}^{n} {Q})^{k_{r-1}-1} \right> }\, .
\end{eqnarray}

As a consequence of the proposition, we have a closed formula for the
two point functions on $G_r (C^n)$, containing $x_1$ and $x_r$ given
by
\begin{equation}
{\ds \left<x_1^{r(n-r)-rk_{r-1}} x_r^{k_{r-1}} \right>}=
{\ds (r(n-k_{r-1} -r))!\prod_{\ell =0}^{r-1} \frac{\ell
!}{(n-k_{r-1}-1 +\ell)!} }\,,
\end{equation}
which is obtained from the self-intersection formula equation (12)
simply by the shift $n\rightarrow n-k_{r-1}$. In particular, on $G_2
(C^n)$, we have the following closed formula
\begin{eqnarray} 
{\ds \left<x_1^{2n-4-2k} x_2^k \right> }= {\ds \frac{(2(n-2-k))!}
{(n-k-2)!(n-k-1)!} } ,
\end{eqnarray}
This particular case, that we denote by $I_k^n$, was also obtained
using topological Kazama-Suzuki models based on complex Grassmannian
\cite{B}. If we set $k_{r-1}=n-r$ in equation (16) one obtains $\left<
x_r^{n-r} \right>=\int_{G_r(C^n)} x_r^{n-1} =1$ as was shown in \cite
{Wi}. The closed formula given by equation (16) is consistent with the
proposition above, since the formula is automatically invariant under
the shifts $n \rightarrow n+1$ and $ k_r \rightarrow k_r +1$, which in
turn gives the two point function on $G_r (C^{n+1})$.
 
\section{Intersection numbers on the space of holomorphic maps to a
Grassmannian} 

Here and in the next section we will give two explicit formulas 
for the intersection numbers
on the space of holomorphic maps of degree $d$ from $CP^1$ into
$G_r(C^n)$ for $r=2,3$. This space of maps is denoted by $Hol_d (CP^1
\rightarrow G_r (C^n))$, the space of instantons of degree $d$. The
intersection numbers on $Hol_d (CP^1 \rightarrow G_r (C^n))$ will be
computed using the deformed potential $\widetilde {W}_{n+1} (x_1
,\dots , x_r)=W_{n+1}(x_1 , \dots ,x_r) +(-1)^r q x_1$ , that
reproduces the quantum cohomology $H_q^* (G_r(C^n ),C)=$ $ \frac{C[x_1
, \dots x_r , q]} {({ \partial \widetilde {W}_{n+1}}/{ \partial x_1},
\dots, { \partial \widetilde {W}_{n+1}}/{\partial x_r})}$ \cite{Wi}.
This means that we will use formally the same formula for the
intersections on the Grassmannian carried out in the last section,
however, the objects inserted in the correlators are the pullbacks of
the cohomology classes (Chern classes) to the parametrizing space of
holomorphic maps of degree $d$. These will be denoted again by $x_i$
$(1 \leq i \leq r)$ such that the total power of the product of these
classes is the dimension of $Hol_d (CP^1 \rightarrow G_r (C^n))$,
which is, $r(n-r)+nd$ \cite{St}. We will see that the intersection
numbers on $Hol_d (CP^1 \rightarrow G_2 (C^n))$ generates alternating
Fibonacci numbers for $n=5$, the Pell numbers for $n=6$, and for $n
\geq 7$ the intersection numbers generates a random walk of a particle
on a line with absorbing bariers \cite{Sl,E}. The self-intersection
formula for $\left<x_1^{2(n-2)+nd}\right>$, which is a special case of
our two-point function given below on $Hol_d (CP^1 \rightarrow G_2
(C^n))$ agrees with that computed in \cite {In} for $n=5$. We have
also checked the geometrical meaning of the quantum correction
\cite{R} associated with the topological $\sigma $-model on $CP^1$
with values in the Grassmannian $G_r(C^n)$, in which computing
correlators on $Hol_d (CP^1 \rightarrow G_r (C^n))$ is equivalent to
doing computations on $Hol_{d-1} (CP^1 \rightarrow G_r (C^n))$
provided we set $x_ry_{n-r} =1$.

In the following, we first write the correlators on $Hol_d(CP^1
\rightarrow G_r(C^n))$ in terms of the Chern roots $q_i$, 
\cite{Wi,Ber,Ra} then we will compute explicitly the formula for 
the intersection numbers for $r=2,3$. The computations on $Hol_d(CP^1
\rightarrow G_3 (C^n))$ are lengthy we will only give the final
formula. The correlators on $Hol_d (CP^1 \rightarrow G_r (C^n))$ are 
given by
\begin{equation}
{\ds \left<x_1^{r(n-r)+nd-2k_1 -\cdots rk_{r-1}} x_2^{k_1}\cdots 
x_r^{k_{r-1}}\right>}
= 
{\ds (-1)^{\frac{r(r-1)}{2}}\!\!\!\!\!\!\sum_{d \widetilde {W}_{n+1} 
=0}\!\!\!\!\!\!\frac{x_1^{r(k-r)+nd-2k_{1} - \cdots -rk_{r-1}} 
x_2^{k_1}\cdots x_r^{k_{r-1}}}{h}}   
\end{equation}
where the summation is over a finite number of critical points of 
$\widetilde {W}_{n+1}(x_1, \cdots x_r)$ and $h= \det ( \partial_i 
\partial_j \widetilde {W})$. In terms of the Chern roots $q_i$, the
potential is given by   
\begin{equation} 
{\ds \widetilde {W}_{n+1} (q_i) }= {\ds \sum_{i=1}^{r} 
\frac {q_i^{n+1}}{n+1} + (-1)^r q_i } .
\end{equation} 
The hessian in terms of the $q_i 's$ on the critical points \cite{Wi} 
is 
\begin{equation}
{\ds  det\ \left[ \frac{ \partial {\widetilde {W}_{n+1}}}{ \partial 
q_i \partial q_j} \right]_{d \widetilde {w}_{n+1} =0} } = {\ds det\  
\left[ \frac{\partial^2 {\widetilde {W}}}{ \partial x_i \partial x_j}
\right] \Delta^2 },  
\end{equation}
where $\Delta=\prod_{j<k}(q_j-q_k)$ is the Vandermond determinant
which is the Jacobian for the change of variables from $q_i$ to
$x_j$. Therefore, the hessian in terms of the Chern roots is given by 
\begin{equation} 
{\ds h(q_1, \cdots ,q_r) }= {\ds det\ \left[ \frac{ \partial^2 
\widetilde{W} }{\partial {x_i} \partial {x_j}} \right] }
={\ds \frac{n^r (q_1, \cdots , q_r)^{n-1}}{ \Delta ^2} } . 
\end{equation} 
Since the Vandermonde determinant vanishes for $q_i = q_j$, the
summation over the critical points given by equation (18) involves
only distinct roots $q_i (1 \leq i \leq r )$ of the polynomial of
degree $n$, of the form $d \widetilde {W}_{n+1} = x^n +(-1)^r$ and
hence the product of the roots satisfy the identity $(q_1 \cdots
q_r)^n =1$. By using the facts $q_i^n = -1,i=1,2$, $x_1 =q_1 + q_2$,
$x_2 =q_1 q_2$ for $r=2$ and making the change of variables $q_i =
\omega \xi_i $ with $\omega^n =-1, \xi_i^n =1$, the two-point
functions on $Hol_d (CP^1 \rightarrow G_2 (C^n))$, that we denote 
by $I_k^{n,d}$, in terms of the new variables $\xi_i$ are
\begin{eqnarray} 
{\ds \left<x_1^{2(n-2)+nd-2k} x_2^k \right> }
&=& {\ds - \sum_{{d \widetilde {W}=0}}
\frac{x_1^{2(n-2)+nd-2k} x_2^k}{h} } \nonumber \\
&=& {\ds \frac{(-1)^{d+1}}{2 n^2} \sum_{{ \xi_i^n =1, 
\xi_1\neq\xi_2}}\left[(\xi_1+\xi_2)^2-4\xi_1\xi_2\right]}
%\nonumber \\ {}\nonumber \\  
{\ds ( \xi_1 + \xi_2 )^{2(n-2) +nd -2k} ( \xi_1 \xi_2 )^{k+1}},
\nonumber\\
\end{eqnarray}
where a factor $1/2$ was inserted in order to avoid overcounting,
since the $x_i 's$ are symmetric in the $q_i 's$. The restriction $
\xi_1 \neq \xi_2$ can be lifted provided we subtract from the sum
terms with $ \xi_1 = \xi_2$.  In our case these terms do not
contribute, therefore, we obtain
\begin{eqnarray}
{\ds \left< x_1^{2(n-2)+nd-2k} x_2^k \right> }=
{\ds  \frac{(-1)^{d+1}}{2} \frac{1}{n^2} \sum_{{\xi_i^n =1}} 
 \left[ ( \xi_1 + \xi_2)^2 -4 \xi_1 \xi_2 \right] } 
\times &&\nonumber \\ 
{\ds ( \xi_1 + \xi_2 )^{2(n-2) +nd-2k} ( \xi_1 \xi_2 )^{k+1}}&&. 
\end{eqnarray} 
If we set $z= \xi_1 \xi_2^{-1}$ in equation (23), then the above 
summation will be over a single $n$-th root of unity $z$, {\it i.e.}, 
\begin{eqnarray}
{\ds \left< x_1^{2(n-2)+nd-2k} x_2^k \right> }= 
{\ds \frac{(-1)^{d+1}}{2} \frac{1}{n^2} \sum_{{z^n}=1} 
\left[ (1 + z)^2-4 z\right]\left(1+z\right)^{2(n-2)+nd-2k}\left(z
\right)^{k+1} }\qquad\qquad&&
\nonumber \\ 
={\ds\frac{(-1)^{d+1}}{2}\frac{1}{n}\sum_{{z^n=1}}\left[\sum_{
{\ell\geq 0}}\binom{2n - 2 + nd -2k}{ \ell } z^{ \ell + k +1} 
- 4 \sum_{{\ell^\prime \geq 0}}\binom{2n - 4 + nd -2k}{ \ell^\prime} 
z^{ \ell^\prime + k +2} \right] }.&&
\end{eqnarray}
The summations of $z^{ \ell + k+ 1}$, $z^{ \ell^\prime +k +2}$ over
the $n$-th roots of unity are non-vanishing only if \footnote{we have
used the identity, $\sum_{z^n =1} z^r= n$, if $r\equiv 0\, mod\,(n)$, 
and vanishing otherwise.} $\ell+k+1=nq$, $\ell^\prime+k+2=
nq^{\prime}$. Finally, explicit computation yields: 
\begin{equation}
{\ds \left< x_1^{2(n-2)+nd-2k} x_2^k \right> } = 
{\ds \frac{(-1)^{d+1}}{2} \sum_{{q \in \left\{ 1,2,\cdots \right\}}} 
\left[ \binom{2n - 2 + nd -2k}{ qn -(k+1)} -4 \binom{2n - 4 + nd -2k}
{qn -(k+2)}\right] }.
\end{equation}
If we set $k=0$ in the above formula then we obtain the explicit
formula for the self-intersection on $Hol_d (CP^1 \rightarrow G_2
(C^n))$, on the other hand, setting $d=0$ gives the two point
functions on $G_2 (C^n)$ obtained in the previous section, equation
(17). Using conformal field theory an expression for the
self-intersection on $Hol_d ( \Sigma_g \rightarrow G_2 (C^5))$ was
obtained \cite{In} where $\Sigma_g$ is a Riemann surface of genus
$g$. When the genus $g=0$, this formula can be written as follows:
\begin{eqnarray} 
{\ds F(0,d)}&=&{\ds\frac{1}{\sqrt {5}}\left[ \left( \frac{\sqrt {5}+1}
{2} \right)^{5(d+1)}+(-1)^d \left( \frac{\sqrt {5}-1}{2} \right)^{5(d+1)}\right]} 
\nonumber \\ 
&=&{\ds\frac{1}{\sqrt {5}}\left[ \left( \frac{\sqrt{5}+1}{2} \right)^{5(d+1)} 
-\left( \frac{1-\sqrt {5}}{2} \right)^{5(d+1)}\right]},
\end{eqnarray}
which is the well known Binet's formula for the Fibonacci numbers
$F_{5(d+1)}$, \cite {Ri}. We have checked for many values of $d$ that
this formula agrees with ours, and therefore we should have the
following mathematical identity on $Hol_d (CP^{ \prime}\rightarrow G_2
(C^5))$:
\begin{equation}
{\ds\left<x_1^{6+5d}\right>}={\ds\frac{(-1)^{d+1}}{2}\sum_{q\in\left\{ 
1,2,\cdots\right\}}\left[\binom{8+5d}{5q-1}-4\binom{6+5d}{5q-2}\right]}
= F_{5(d+1)}\,.
\end{equation}
By an explicit computation for the two point functions on $Hol_d (CP^1
\rightarrow G_2 (C^5))$ see table 3 for various $k$ and for fixed
$d$, one can see that we should have the identity
\begin{equation}
\left< x_1^{6+5d-2k} x_2^k \right> = F_{5(d+1) -2k} . 
\end{equation}
The intersection numbers given by $F_{5(d+1)-2k}$ corresponds to the
alternate Fibonacci numbers for $ 0 \leq k \leq [5(d+1)/2]$ with $d$
fixed, and for $k=[5(d+1)/2]$ the intersection numbers are equal to
one or zero depending on whether the degree of the holomorphic maps $d$
is even or odd.  This seems to hold for every $n$. For $n=4$, the
intersection numbers $Hol_d (CP^1 \rightarrow G_2(C^4))$ are powers of
2, as one can see from the table 3. For $n=6$, we obtain two
possible sequences of Pell numbers \cite{Sl}, when $d$ is odd the
general term is $\frac{3^m -1}{2}$, and the other sequence given by
$\frac{3^m +1}{2}$ for even degree. In general, for $n \geq 7$, the
intersection numbers generate a random walk with absorbing barriers
\cite{Sl,C}.  This is a one dimensional random walk, in which the
particle starts at point $1$ and arrives eventually at the point $n$,
the particle may never visit $0$, i.e. the points $0$ and $n$ are
absorbing barriers this happens when the degree is odd. When the
degree $d$ is even, the intersectoin numbers generate a random walk
on a line for a particle that starts at point $n-1$ see table 4.

\section{The Correlators on $Hol_d (CP^1 \rightarrow G_3
(C^n))$} 

In computing all the correlators on $ Hol_d (CP^1 \rightarrow
G_3(C^n))$ that we denote by $I_{k_1,k_2}^{n,d}$we follow the same
technique as for the two-point functions computed in section 3.  Using
equations (18), (21) and after some algebra, the correlators can be written
as
\begin{eqnarray} 
&&{\ds\left<x_1^{3(n-3)+nd -2k_1 -3k_2} x_2^{k_1} x_3^{k_2}\right>}
={\ds -\frac{1}{n^3} \sum_{{q_i^n=1,i=1,2,3}}(q_1 -q_2)^2 
(q_2 -q_3)^2 (q_1-q_3)^2}\times 
\nonumber \\{}\nonumber\\ 
&&\qquad\qquad\qquad
{\ds (q_1 +q_2 +q_3 )^{3(n-3)+nd -2k_1-3k_2} 
(q_1 q_2 +q_1 q_3 +q_2 q_3)^{k_1} (q_1 q_2 q_3 )^{k_2 +1} } 
\nonumber \\{}\nonumber\\  
&&\qquad\qquad\qquad
= {\ds - \frac{1}{6}\sum_{{p,q\in\left\{1,2,\cdots \right\}}}
\sum_{{s,t=0}}^{2} (-1)^{s+t}\binom{2}{s} \binom{2}{t} 
\sum_{{\ell^{ \prime}=0}}^{k_1} \sum_{
{ \ell^{\prime\prime}=0}}^{\ell^{ \prime}} 
\binom{k_1}{ \ell^{ \prime}} \binom{ \ell^{ \prime}}
{ \ell^{\prime \prime}} }\times
\nonumber  \\ {}\nonumber\\ 
&&\qquad\qquad\qquad
{\ds \frac{x!}{(x-y)!} \left[\frac{1}{(z)!(w)!} +\frac{1}{(z+2)!}
-\frac{2}{(z+1)(w-1)!} \right] }\,, 
\end{eqnarray}
where 
\begin{eqnarray}
w&=&qn-( \ell^{\prime} +k_2 +t+1)\,,
\nonumber \\
x&=&3(n-3)+nd -2k_1 -3k_2\,, 
\nonumber \\
y&=&(p+q)n -( \ell^{\prime} + \ell^{\prime \prime}
+k_1 +2k_2 +4+s+t)\,, 
\nonumber \\
z&=&pn-(k_1 +k_2 +s+3- \ell^{\prime \prime})\,.
\end{eqnarray} 
If we set $d=0$, $k_1=k_2=0$ in the above formula, then we obtain the
number  
\begin{eqnarray}
I^{n,0}_{0,0}=\frac{2(3(n-3))!}{(n-3)!(n-2)(n-1)!} , \, \nonumber 
\end{eqnarray}
which is the self-intersection formula for $<x_1^{3(n-3)}>$ on $G_3
(C^n)$ given by equation (12). We also have checked that the above
formula for $d=0$ gives the intersections numbers on $G_3(C^5)$ and
$G_3(C^6)$ and therfore setting $d=0$ in equation (29) we obtain the
formula for the intersection numbers $I_{k_1,k_2}^n$, on $G_3 (C^n)$.

Let us check the implication of the the geometrical meaning of the
quantum correction \cite{R} using our formula, equation (29). As was
mentioned in the beginning of this section, the quantum correction
implies that correlators on $Hol_d(CP^1 \rightarrow G_3(C^n))$ are
identical to those on $Hol_{d-1}(CP^1 \rightarrow G_3(C^n))$ provided
we set $x_3 y_{n-3}=1$. This can be seen by considering the following
simple example: suppose we want to evaluate the correlator $\left<
x_1^7 x_2 x_3 y_3 \right>_{d=1}$, where $y_3= x_1^3 -2x_1x_2 +
x_3$. Then using the results indicated in table 5, where
$I_{1,1}^{6,1} =171$, $I_{2,1}^{6,1} =86$ and $I_{1,2}^{6,1} 
=22$ we have $\left< x_1^7 x_2 x_3 y_3 \right>_{d=1}=21$ which is $
\left< x_1^7 x_2 \right>_{d=0} = I_{1,0}^{6}$ see table 2.

\section{Conclusion} 

In section two, we obtained a closed formula for the
two-point functions $\left< x_1^{r(n-r)-rk_{r-1}} x_r^{k_{r-1}}
\right>$ on $G_r(C^n)$ given by equation (16). When we set $r=2$,
two-point functions on $G_2(C^n)$ are obtained. Also, all the
correlators $\left< x_1^{3(n-r)-2k_1-3k_2} x_2^{k_1} x_3^{k_2}
\right>$ on $G_3(C^n)$ are obtained by restricting our formula on the
space of holomorphic maps of degree $d$ to constant maps, i.e., $d=0$.
The closed formulas, obtained here are extensions of the Schubert
formula equation (12) that computes the self-intersection $\left<
x_1^{r(n-r)} \right>$.

In section three, we obtained an explicit formula for the two-point
functions on the space of holomorphic maps of degree $d$ from $CP^1$
into $G_2(C^n)$. This formula generates well known numbers like the
Fibonacci numbers for $n=5$ and the Pell numbers for $n=6$ \cite
{Sl}. However, when $n \geq 7$ the formula generates a random walk of
a particle on line with absorbing barriers \cite{Sl,C} that starts at
the point $1$ and eventually reaches the point $n$, if the $d$ is odd.
When $d$ is even the particle starts at the point $n-1$ see table 4
and \cite {E}. At the moment, we do not understand this connection. It
would be nice if this can be understood from both mathematics and
physics.

Also on the moduli space of holomorphic maps into $G_r(C^n)$, we have
computed intersection numbers using equation (29), these numbers
follow well organized patterns for given $n$. For $n=5$, the
intersection numbers are given by the Fibonacci numbers, this is
expected since $G_3(C^5)$ and $G_2(C^5)$ are dual to each
other. Setting $n=6$, one obtains a pattern like that in tables 5 and
6 for any degree $d$. For $n=7$, we have found sequences of numbers
such that the ratio of any consecutive numbers behave like that of
$L_n/F_n$, the $n^{th}$ Lucas number by the $n^{th}$ Fibonacci number,
which is known to be $\sqrt {5}$.
\par\vspace{.8cm}
{\Large\bf Acknowledgement}
\vspace{.6cm}\par
I would like to thank Z. Badirkhan and S. F. Hassan for their help
during this work. I would also like to thank M. S. Narasimhan,
N. Nitsure and C. Reina for discussions and the ICTP and SISSA for
support and hospitality.

\newpage
\begin{table}
 \begin{center}
  \begin{tabular}{|l l| l | l | l | l | l | l|}
   \hline
      &k & 0   & 1   & 2  & 3  & 4  & 5 \\
   n  &  &     &     &    &    &    &    \\ \hline
   4  &  & 2   & 1   & 1  &    &    &    \\   
   5  &  & 5   & 2   & 1  & 1  &    &    \\   
   6  &  & 14  & 5   & 2  & 1  & 1  &    \\   
   7  &  & 42  & 14  & 5  & 2  & 1  & 1   \\ \hline
  \end{tabular}
  \caption{ Intersection numbers $I_k^n$ on $G_2 (C^n)$ for $n=4,5,6,7$}
  \end{center}
\end{table}
\begin{table}
 \begin{center}
  \begin{tabular}{| l | l | l | l | l | l|}
   \hline 
          &     &        &     &      &      \\
   $(k_1 ,k_2)$ & $I_{k_1 ,k_2}^5$ & $(k_1 ,k_2)$ & $I_{k_1 ,k_2}^6$ &
$(k_1 ,k_2)$ 
   & $I_{k_1 ,k_2}^7$ \\ \hline
   (0,0)  & 5  &  (0,0)  & 42 & (0,0) & 462 \\
   (1,0)  & 3  &  (1,0)  & 21 & (1,0) & 210 \\
   (2,0)  & 2  &  (2,0)  & 11 & (2,0) & 98  \\
   (3,0)  & 1  &  (3,0)  & 6  & (3,0) & 47  \\
   (1,1)  & 1  &  (0,1)  & 5  & (0,1) & 42  \\
   (0,1)  & 1  &  (4,0)  & 3  & (4,0) & 23  \\
   (0,2)  & 1  &  (1,1)  & 3  & (1,1) & 21  \\
          &    &  (2,1)  & 2  & (5,0) & 11  \\
          &    &  (0,2)  & 1  & (2,1) & 11  \\
          &    &  (1,2)  & 1  & (3,1) & 6   \\
          &    &  (0,3)  & 1  & (6,0) & 5   \\
          &    &         &    & (0,2) & 5   \\
          &    &         &    & (4,1) & 3   \\
          &    &         &    & (1,2) & 3   \\
          &    &         &    & (2,2) & 2   \\
          &    &         &    & (0,3) & 1   \\
          &    &         &    & (0,4) & 1   \\
          &    &         &    & (1,3) & 1   \\
          &    &         &    & (3,2) & 1    \\  \hline
   \end{tabular}
   \caption{ Intersection numbers $I_{k_1 k_2}^n$ on $G_3 (C^n)$ for
$n=5,6,7$} 
   \end{center}
\end{table}
\begin{table}
 \begin{center} 
  \begin{tabular}{| l | l | l || l | l || l | l |}
   \hline 
   &       &    &     &      &      &        \\
 k & $I_k^{4,3}$ & $I_k^{4,4}$ & $I_k^{5,2}$ & $I_k^{5,3}$ &
$I_k^{6,2}$  &$I_k^{6,3}$  \\ \hline
0   & 128 & 512 & 610 & 6765 & 9842 & 265720  \\
1   & 64  & 256 & 233 & 2584 & 3281 & 88573    \\
2   & 32  & 128 & 89  & 987  & 1094 & 29524    \\
3   & 16  & 64  & 34  & 377  & 365  & 9841     \\
4   & 8   & 32  & 13  & 144  & 122  & 3280     \\
5   & 4   & 16  & 5   & 55   & 41   & 1093     \\
6   & 2   & 8   & 2   & 21   & 14   & 364      \\
7   & 1   & 4   & 1   & 8    & 5    & 121      \\
8   & 0   & 2   & 1   & 3    & 2    & 40       \\
9   &     & 1   &     & 1    & 1    & 13       \\
10  &     & 1   &     & 0    & 1    & 4         \\
11  &     &     &     &      &      & 1         \\
12  &     &     &     &      &      & 0         \\ \hline
 \end{tabular}
  \caption{Intersection numbers $I_k^{n,d}$ for $n=4, d=3,4$; $n=5,
d=2,3$ and $n=6, d=2,3$}
   \end{center}
\end{table}
\begin{table}
 \begin{center}
  \begin{tabular}{| l | l | l | l | l | l |}
   \hline 
    &      &        &       &        &          \\
 k & $I_k^{7,1}$ & $I_k^{7,2}$ & $I_k^{8,1}$ & $I_k^{9,1}$ &
$I_k^{10,1}$ \\ \hline 
0   & 2380 & 147798 & 15504 & 100947 & 657800   \\
1   & 728  & 45542  & 4488  & 28101  & 177859   \\
2   & 221  & 14041  & 1288  & 7752   & 47562     \\
3   & 66   & 4334   & 364   & 2108   & 12597      \\
4   & 19   & 1341   & 100   & 560    & 3264       \\
5   & 5    & 413    & 26    & 143    & 820       \\
6   & 1    & 131    & 6     & 34     & 196       \\
7   & 0    & 42     & 1     & 7      & 43        \\
8   &      & 14     & 0     & 1      & 8          \\
9   &      & 5      &       & 0      & 1           \\
10  &      & 2      &       &        & 0           \\
11  &      & 1      &        &       &          \\
12  &      & 0      &        &       &     \\ \hline
 \end{tabular}
  \caption{Intersection numbers $I_k^{n,d}$ for $n=7, d=1,2$ and
$n=8,9,10, d=1$ } 
   \end{center}
\end{table}
\begin{table}
 \begin{center}
  \begin{tabular}{| l  l | l | l | l | l | l | }
   \hline
         & $k_2$ &  0     &  1    &  2    & 3    &  4  \\ 
   $k_1$ &       &        &       &       &      &     \\ \hline 
      0  &       & 2730   & 341   & 43    &  6   & 1   \\
      1  &       & 1365   & 171   & 22    &  3   & 0    \\
      2  &       &  683   & 86    & 11    &  1   & 0    \\
      3  &       &  342   & 43    & 5     &  0   &      \\
      4  &       &  171   & 21    & 2     &      &      \\
      5  &       &  85    & 10    &       &      &      \\  
      6  &       &  42    & 5     &       &      &      \\
      7  &       &  21    &       &       &      &      \\
         &       &         &      &       &      &     \\ \hline
   \end{tabular}
   \caption{Intersection numbers $I_{k_1,k_2}^{6,1}$}
   \end{center}
\end{table}
\begin{table}
 \begin{center}
  \begin{tabular}{| l  l | l | l | l | l | l | l | }
   \hline 
       & $k_2$&   0   &   1  &   2   &   3   &  4    &  5  \\
 $k_1$ &      &       &      &       &       &       &     \\ \hline
    0  &      & 17476 & 21845 & 2731 &  342  &  43   &  5   \\
    1  &      & 87381 & 10923 & 1366 &  171  &  21   &  2   \\    
    2  &      & 43691 & 5462  & 683  &  85   & 10    &  1    \\
    3  &      & 21846 & 2731  & 341  &  42   & 5     & 1     \\
    4  &      & 10923 & 1365  & 170  &  21   & 3     &       \\
    5  &      & 5461  & 682   & 85   &  11   &       &       \\
    6  &      & 2730  & 341   & 43   &  6    &       &       \\
    7  &      & 1365  & 171   & 22   &       &       &        \\
    8  &      & 683   & 86    &      &       &       &        \\
    9  &      & 342   & 43    &      &       &       &        \\
    10 &      & 171     &       &    &       &       &        \\ \hline
   \end{tabular}
   \caption{ Intersection numbers $I_{k_1 k_2}^{6,2}$ }
   \end{center}
\end{table}

\begin{thebibliography}{99} 
\bibitem{H} W.V.D. Hodge and D. Pedoe {\it Methods of Algebraic
Geometry} {\bf 2}, Cambridge Univ. Press, (1952).
\bibitem{Gr} P. Griffiths and J. Harris, {\it principle of Algebraic
Geometry}, J. Wiley (1978).
\bibitem{F} W. Fulton {\it Intersection Theory}, Springer -Verlag
Berlin Heidelberg (1984).
\bibitem{V} C. Vafa  {\em Mod. Phys. Lett.} {\bf A6}, 337 (1991). 
\bibitem{C} N. Chair {\it Grassmannian Cohomology Rings and Fusion
Rings from Algebraic Equations} {\em hep-th/9704138}
\bibitem{In} K. Intriligator {\em Mod. Phys. Lett.} {\bf A6} 3543
(1991). 
\bibitem{V3} C. Vafa {\it Topological Mirrors and Quantum Rings}, {\em
in Essaya on Mirrors Manifolds, ed. S-T. Yau} (International Press,
1992). 
\bibitem{W2} E. Witten {\em Nucl. Phys.} {\bf B340}, 281 (1990). 
\bibitem{Wi} E. Witten, {\it The Verlinde Algebra and the Cohomology
of the Grassmannian}, preprint IASSNS-HEP 93/41, December (1993). 
\bibitem{Sl} N.J.A. Sloane and Simon Plouffe, {\it The Encyclopidia of
Integer Sequences} Academic Press (1995).
\bibitem{E} C. J. Everrett and P.R. Stein {\em Discrete Mathematics}
{\bf 17} 27 (1977).
\bibitem{V2} W. Lerche, C. Vafa and N. Warner,{\em Nuc. Phys.} {\bf
B324} 427-474 (1989).
\bibitem{G} D. Gepner, {\em Comm. Math. Phys.} {\bf 141}  381-411
(1991). 
\bibitem{BT} R. Bott and L. W. Tu, {\it Differential Forms in
Algebraic Topology}, Springer (1982).
\bibitem{B}  M. Blau, F. Hussein, G. Thompson {\em Nucl. Phys.} {\bf
B488} 599 (1997).
\bibitem{St} S. Stromme {\it On Parametrized Rational Curves in
Grassmannian Varieties} In space curve eds. F. Ghione et al. Springer
-Verlag Lecture Notes in Mathematics (springer, 1987).
\bibitem{R} D. Franco and C. Reina {\em Comm. Math. Phys.} {\bf 174}
137 (1995). 
\bibitem{Ber} A. Bertram, G. Daskalopoulos and R. Wentworth, {\it
Gromov Invariant for Holomorphic Maps from Riemann Surfaces to a
Grassmannian}, Preprint (April, 1993).
\bibitem{Ra} M.S. Ravi, J.Rosenthal and X. Wang, {\it Degree of the
Generalized Plucker Embeding of a Quote Scheme and Quantum Cohomology}
Preprint (1993). 
\bibitem{Ri} T.J. Rivlin, {\it The Chebyshev Polynomials}, $2^{nd}$
{\em edition - 
J. Wiley} New York (1990).
\end{thebibliography}
\end{document}